Manuscript

# Spectrum accessing optimization in congestion times in radio cognitive networks based on chaotic neural networks

*Mahdi Mir\*,*
*Department of Electrical Engineering, Ferdowsi University of Mashhad, Mashhad, Iran*

**Abstract**

Based on the theory of the Federal Communications Commission, the spectrum available on cognitive radio networks is limit and the non-optimal use of the spectrum necessitates the need for a telecommunications model, so that this pattern can exploit the existing spectral positions. In this spectrum subscription scenario, when the primary users are not present, it is also possible to assign this telecommunication to tenants who are unauthorized or secondary. The challenge of using this scenario is to allocate time-frequency resources to them and how to access nodes in one channel without any interactions between primary and secondary users and the throughput will increase. The main idea of this research is using chaotic recurrent neural network for improving access to spectrum in congestion times and the main purposes are reduce interference and increase throughput in cognitive radio networks. In this method, in addition to the throughput, the amount of unwanted blockage of packets, the reduction of the cost of operations for secondary users, the hardware requirements for secondary users and the coefficient of justice are considered which in fact, it is a new channel assignment process with respect to the environment response, the updates the probability that the channels are empty in subsequent periods, and increases the permeability by reducing interference with chaotic recurrent neural network.

**Keywords**: cognitive radio network, spectrum access, primary user, secondary user, chaotic recurrent neural network

## 1- Introduction

Cognitive Radio Network is a solution for spectrum congestion problem which use opportunistic frequency band which is not completely occupied by licensed users. Cognitive radio network define as Federal Communication Commission definition: a radio or system that sense electromagnetism operation and can automatically adjust its radio parameters to optimize system operations such as maximize operational capacity and reduce interference [1].

---

\* Corresponding Author.
E-mail address: Mahdimir.ir@gmail.com (Mahdi Mir)

In the term of cognitive radio network, primary users defined as a group which have the higher priority or legal right to use a particular part of the spectrum. On the other hand, secondary users have lower priority and must use the spectrum in a way that does not interfere with the primary users. So, secondary users must have abilities such as spectral measurement to recognize the existence of the main use. The primary user does not have concerns about the behavior of the smart grid and there is no need for special ability to coexist with it. Secondary users which have not license and should not interfere with primary user transfers, so as the soon as primary users identified, secondary user must react with changing radio frequency power and channel rate. It's because of their transfers should not lower the quality of the service of the primary users. In addition, secondary users must coordinate their access to the existing channel and avoid collisions between smart users [1].

2- Recent Researches
Various studies proposed in spectral and channel accessing optimization for primary and secondary users in cognitive radio networks. In [2], artificial intelligent methods used for intelligence and cognitive radio networks design. In [3], Bayesian learning used for predicting availability or not of channels which the learning process is simplified considering the geometric distribution for the channel model. In [4], MAC cognitive inferiority protocol proposed which allow secondary users that independently specified without the use of a central coordinator or telecommunications channel for finding spectral opportunities based on Markov Chain Decision theory.

In [5, 6], investigate about routing protocols and channel assignment methods of cognitive radio mobile ad hoc networks. First, a subset of candidate channels that can be used as a secondary network channel are then determined and then explained about when they will be placed on the candidate channels. This method is called moving the informed channel [7-9].

In [10] proposed a method for grid flow for selecting network to survey about secondary users in cognitive radio networks. In [11], proposed a method for preventing overlap of primary and secondary user cooperation with existing rules of cognitive radio network. They have been able to create rules among users to enhance the participation of two users together to transfer data from the user's primary side by maximizing the space which measure spectrum with high security in online environment.

In [12] proposed a method for power control for primary and secondary user in cognitive radio network which modeled a new cost function for transferring power and sum of them created Nash equilibrium. Including channel selection methods and spectrum detection that has been investigated up to now, can be notice to the classic test is the likelihood ratio [13, 14], energy detection [15, 16], detection with adaptive filters such as LMS, NLMS and RMS [17], detection based on spacing signal characteristic [18-20] and other emerging techniques [21].

In [22], a new method for spectrum sensing under primary user emulation attacks simulated in cognitive radio networks. This research used threshold selection to minimize the total error probability. Comparison represented that this method has some advantages in performance with other methods such as conventional techniques. In [23], impact of residual time distributions of spectrum holes on spectrum handoff performance with finite switching delay in cognitive radio networks proposed.

## 3- A review of recurrent neural network

Recurrent neural network works based on reversibility and have hidden neurons. Recurrent neural network have two layers. Hidden and output layer with input layer which has hidden layer too. Both layers of this neural network receive two collection of inputs. First collection have delay of all neurons of outputs (hidden neurons and output) and second collection of inputs contain external input signals. Recurrent neural network with its learning rule have another ability which is real-time learning. Convergence means network learning leads to estimate weights which allow the network to provide the correct output value for each of the input instruction patterns and patterns similar to them. In this learning, Actuator, correct response, optimal input and output, pattern and what types of patterns belong to classes. In these learnings, training errors available; Therefore in practice, it is appropriate to use a learning error to set up network parameters at each step so that, if the same input is applied again, the learning error is less.

## 4- Proposed method

A cognitive radio network is assumed with primary users that set of its channels is in the form of $CH_{Set} = \{1, ..., M\}$ with channel bandwidth like $m \in CH_{set}$ and consider as Hz. Meanwhile, using a system with certain time range consider such as $t \in \{1, ... t\}$. Secondary user select a channel for sensing and resource efficiency. At the end of each time range, the receiver send a message to transmitter for successful transfer operation. When secondary users are more than one in the system, the set of secondary users will be $SU = \{1, ..., N\}$. Channel $m$ with $\theta_m$ probability as free channel and $1 - \theta_m$ probability is busy channel. With considering a random Bernoulli variable in the form of $\zeta_{m(t)}$, it will be characterized that if equal to one, $m$ channel in $t$ time range is free and if equal to zero, other scenarios will occur. It should be noted that secondary user from channel entity vector such as $\theta = (\theta_1, ..., \theta_m)$ do not aware and $C_m(t)$ are the set of secondary users which select $m$ channel in $t$ time range. After occupying the channel by secondary users, $B$ bit sending will possible in $m$ channel at $t$ time range. In this case, there may be three events that describe the equation (1) to (3).

$$if\ \zeta_{m(t)} = 1 \quad |C_{m(t)}| \tag{1}$$

$$if\ \zeta_{m(t)} = 0 \quad |C_{m(t)}| > 1 \tag{2}$$

If equation (1) exist, secondary user will be wait until occupying channel and efficiency of the spectrum up to the next round and if equation (2) exist, secondary users will be overlap to other users and no secondary user can't use selected channel bandwidth over time. Total bits of secondary users in $t$ time range, calculated as equation (3).

$$W_n = \sum_{t=1}^{T} B.\zeta_{m(n,t)}(t).N\ on\ Col_{m(t)}(t) \tag{3}$$

In equation (3), $m(n,t)$ is selected channel in $t$ time period by $n$ secondary user. $N \, on \, Col_{m(t)}$ is random Bernoulli variable that if equaled to one, there is no overlapping and if it equaled to zero, other operation done. The purpose is to maximizing sum of throughputs for all secondary users based on recurrent neural network with training by using equation (4).

$$W = \sum_{t=1}^{T} W_n = \sum_{n=1}^{N} \sum_{t=1}^{T} B . \zeta_{m(n,t)}(t) . N \, on \, Col_{m(t)}(t) \qquad (4)$$

The connectivity is possible from channel $i$ to channel $j$ in $t$ time period and $c$ cost. The total channel connection cost calculated by adding some time range that each secondary user from one channel to other channel connect in $T$ time range. It's needed to use chaotic recurrent neural network for selecting and assigning channels and spectrum in cognitive radio networks. For this purpose, this algorithm is in the form of relations (5) and (6) respectively.

$$P_i(t+1) = P_i(t) + \alpha.(1 - P_i(t))$$
$$P_j(t+1) = P_j(t) - \alpha.P_j(t) \, , \quad \forall j, j \neq i \qquad (5)$$

$$P_i(t+1) = (1-\beta).P_i(t)$$
$$P_j(t+1) = \frac{\beta}{r-1} + (1-\beta).P_j(t), \quad \forall j, j \neq i \qquad (6)$$

Which equation (5) is for simple recurrent neural network and equation (6) is for chaotic model. $\alpha$ and $\beta$ are the learning parameters in recurrent neural network. Proposed method of this research is using chaotic recurrent neural network which this kinds of chaos, will be select $a(t) = a_i$ operation from $\{a_1, \dots, a_r\}$ operation set as an input for cognitive radio network.

Chaotic recurrent neural network algorithm $L_{R-P}$ schema expansion to Q model by introducing various parameters for responding to different environments and situations. According to this description, the response environment for the operation is an element of the set $X = \{X_i^1, X_i^2, \dots, X_i^R, \bar{X}_i^P, \bar{X}_i^{P-1}, \dots, \bar{X}_i^1\}$ which $\{X_i^1, X_i^2, \dots, X_i^r\}$ is data delivery and $\overline{\{X_i^P, \bar{X}_i^{P-1}, \dots, \bar{X}_i^1\}}$ is the set of bit error rate in congestion time of spectral accessing. Corresponding reinforcement signals for X operation set is $\alpha_i^1, \dots, \alpha_i^r$ for data delivery and $\beta_i^1, \dots, \beta_i^r$ for bit error rate. R and P are the number of data delivery and bit error rate. It's assumed that $0 \leq X_i^1 < \dots < X_i^R < m_i < \bar{X}_i^P < \dots < \bar{X}_i^1 \leq 1$ that $m_i$ threshold is considered for reply to data delivery and bit error rate. It is clear that $X_i^1$ is the best mode and $\bar{X}_i^1$ is the worst mode. Using chaotic recurrent neural network for single secondary user scenario for selecting channel and $i$ spectrum assignment for upgrading $P(t+1)$ probability vector is as equation (7).

$$P_j(t+1) = P_j(t) - g_j(P(t))$$
$$P_j(t+1) = P_j(t) - h_j(P(t)) \qquad (7)$$

By noticing to equation (7), it is clear that in first equation, channel $i$ in $t$ time period is free for all $j \neq i$ and in second equation, channel $i$ in $t$ time period is busy for all $j \neq i$. For keeping size of probabilities, it should be $\sum_{j=1}^{M} P_j(t) = 1$ existed. So, equation (8) created.

$$P_i(t+1) = P_i(t) + \sum_{\substack{j=1 \\ j \neq i}}^{r} g_j(P(t))$$

$$P_i(t+1) = P_i(t) + \sum_{\substack{j=1 \\ j \neq i}}^{r} h_j(P(t)) \tag{8}$$

In equation (8) part one, represented that $i$ channel in $t$ time period is free and part two, $i$ channel in $t$ time period is busy. It should be noted that $g_j(\ )$ and $h_j(\ )$ are the package delivery and BER which lead to satisfy equation (9) and they are package delivery and BER function in linear reinforcement schematic such as equation (10).

$$0 < g_j(P) < P_j, 0 < \sum_{\substack{j=1 \\ j \neq i}}^{M}[P_j + h_j(P)] < 1 \tag{9}$$

$$g_j(P(t)) = \alpha P_j(t) \ , \ h_j(P(t)) = \frac{\beta}{M-1} P_j(t) \tag{10}$$

In equation (10), $\alpha$ and $\beta$ ate package delivery parameters and BER and , $0 \leq \beta < 1$ and $0 < \alpha < 1$. Secondary user leads to increase the probability of $i$ selecting channel and reduce the probabilities of other channels to respond to spectrum access. Adoption of dynamic spectrum access control based on the chaotic recursive neural network approach and the updating of this control acceptance is calculated as (11).

$$\begin{cases} \psi(t+1) = \min(\psi(t) + \mu, 1) \\ \psi(t+1) = \max(\psi(t) - \mu, 0) \end{cases} \tag{11}$$

Based on first part of equation (11), secondary user have not overlap in channel selection and spectrometry and second part have it. $\mu$ is the uniform random variable between (0,1) range that occurs by any secondary user when connected to the network; minimum and maximum for $\psi(t)$ boundary is between (0,1). Thus, there is no overlap or interference in improving access to the spectrum during congestion by users.

## 5- Simulation and results

First, it is necessary to mention the variables that have been set up in the cognitive radio network as shown in Table (1).

Table (1) cognitive radio network parameters

| | |
|---|---|
| Channels Number $M$ | 10 |
| Total selected channel in each time range for all users | 20 |
| Time range $t$ | 5 |
| $P_i$ value | 10 |
| $P_j$ value | 10 |
| $\alpha$ value | 0.5 |
| $\beta$ value | 0.5 |
| $r$ value | $\alpha + \beta$ |
| $P$ probability | 0.1 |

| Number of secondary user | 100 |
|---|---|
| Initial throughput value | 3 |
| Size of sending signal $B$ | 100000 bit |
| Iteration of chaotic recurrent neural network | 20 |
| Energy of per bit of data (Jul) | 1 |
| Signal-to-Noise-Ratio (SNR) | (0,9) |

When simulation run ends, total throughput in terms of time range represented in Fig. 1 for the proposed approach, the recurrent neural network, the chaotic model and the classic model.

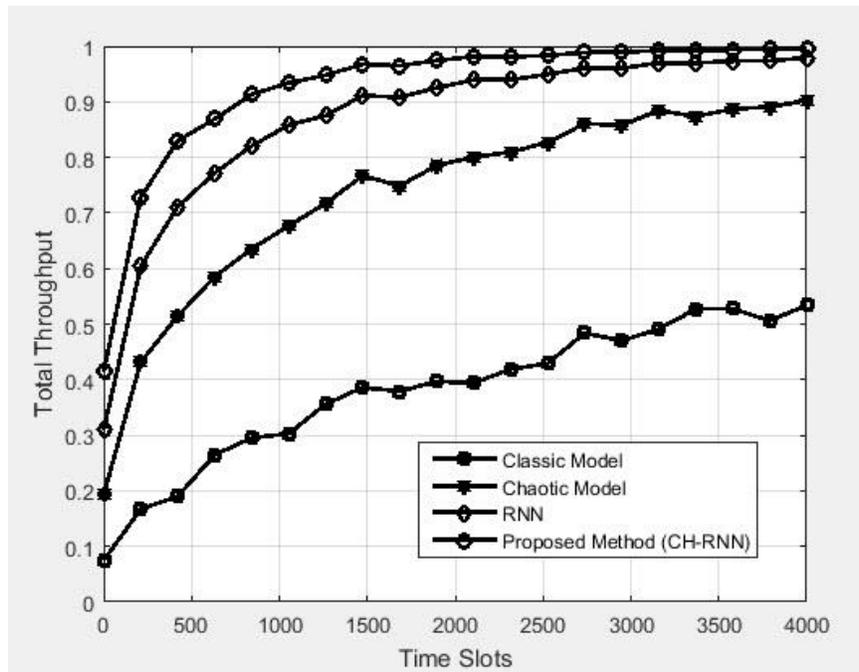

Fig. 1, total throughput in terms of time range

As is clear, proposed method (Ch-RNN) amount of throughput in terms of 4000 ms in 20 iteration is more than any other methods. Also, classic model for accessing the spectrum as well as graphs for the throughput of the time range shown. In fact, a general comparison between the application of various rules of chaotic recurrent neural network for channel and spectrum selection by secondary users in the cognitive radio network has been made. In Fig.2 represented total throughput during congestion access to the spectrum with 100 secondary users.

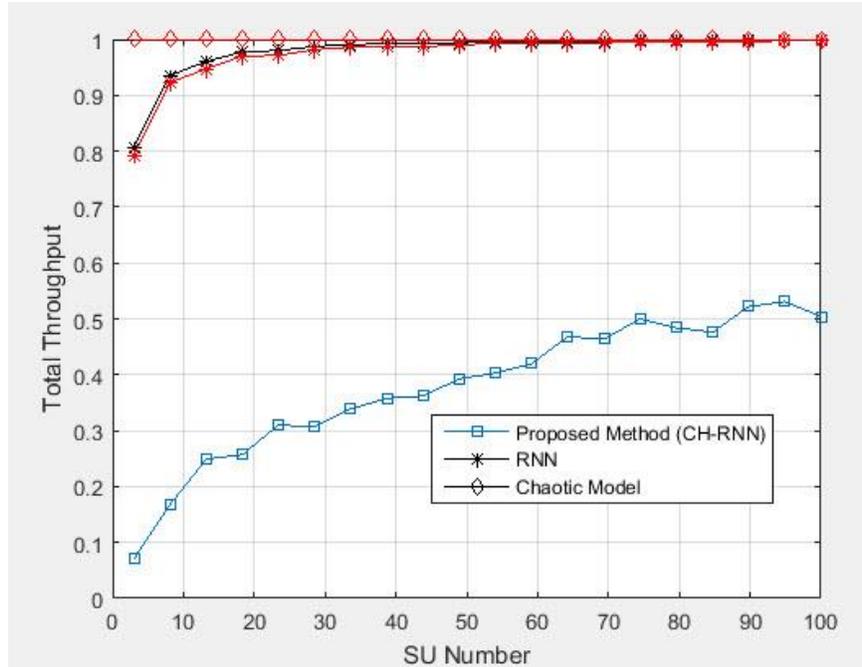

Fig. 2, total throughput to secondary users in congestion access to the spectrum

As can be seen, proposed method has the least amount of throughput in 20 iterations of selecting and assigning channel. But the method of using only a recursive neural network in the red diagram and using only a chaotic model on a black chart has same throughput ratio, but the recurrent neural network in a series of sections shows more throughput. As in Fig. 1, in Fig. 2, a general comparison between the proposed method and the use of only two other models in the cognitive radio network has been performed for selecting the channel and spectrum by secondary users. Subsequently, using the number of bits sent to the network and the signal-to-noise ratio, optimization of channel selection and assignment, as well as acceptance of control in the cognitive radio network, is shown in Fig. 3.

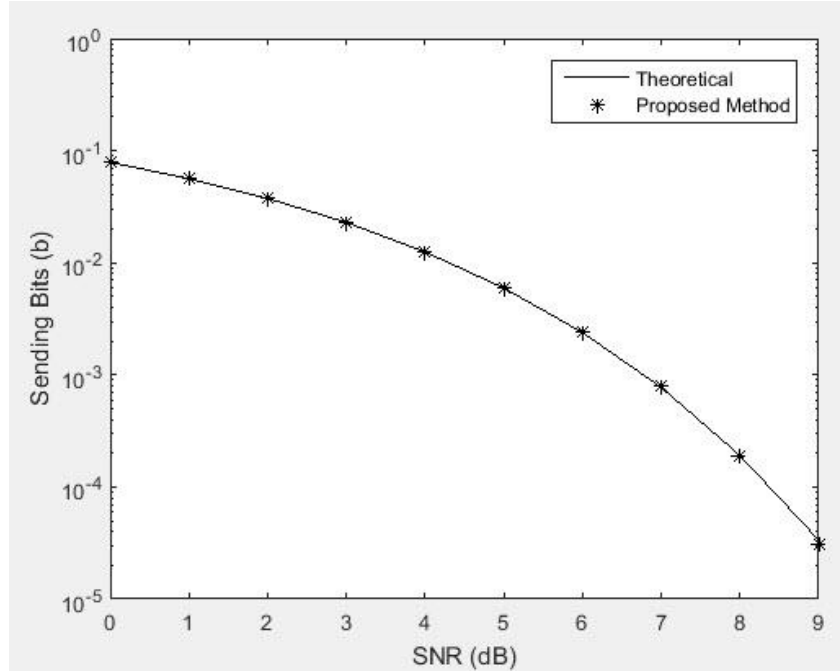

Fig. 3, Channel election and assignment optimization and control acceptance based on the number of bits sent to the network and the amount of signal-to-noise ratio

In the theory of channel selection and assignment, as well as acceptance of control, the optimization has been accomplished by chaotic recurrent neural network approach with points optimized based on the bits being marked. The runtime is also 85.84 seconds.

## 6- Conclusion

Today, due to the increasing use of wireless technologies, one of the most important problems in telecommunication science is the availability of accessible bandwidth constraints. One of the proposed method for using optimized bandwidth usage is channel selection and assignment with spectrum access dynamically. In recent years, cognitive radio network introduces as a tool for optimal use of spectrum. By detecting the surroundings through the sensor, as well as signaling, these radios are able to use secondary sources of radio sources, meaning that they can only use the spectrum when they are not licensed by the user. Secondary users are unlicensed users who should avoid interfering with primary user transfers. When primary users use spectrum, the secondary users are required to dump the channel to prevent the loss of primary users. The proposed method of this research is the secondary user intends to select the channel and the spectrum measurement. In this scenario, chaotic recurrent neural network use which this operation takes place with the acceptance of control in choosing the channel and spectrum. Results represent that proposed method have the better performance in comparison to recent techniques.